\documentstyle[multicol,aps,prb,epsf]{revtex} 

\begin{document}
\author{G. Gomila}
\address{Research Center for BioElectronics and NanoBioScience, \\
Departament d'Electronica, Universitat de Barcelona, \\
C/ Marti i Franques, 1, E-08028 Barcelona, Spain}
\author{T. Gonz\'{a}lez}
\address{Departamento de F\'{\i }sica Aplicada, Universidad de Salamanca, \\
Plaza de la Merced s/n, E-37008 Salamanca, Spain.}
\author{L. Reggiani}
\address{INFM-National Nanotechnology Laboratory,\\
Dipartimento di Ingegneria dell'Innovazione, \\
Universit\`{a} di Lecce, \\
Via Arnesano s/n, I-73100 Lecce, Italy.}
\title{Shot-noise anomalies in nondegenerate elastic diffusive conductors}
\date{\today}
\address{}
\maketitle

\begin{abstract}
We present a theoretical investigation of shot-noise properties in
nondegenerate elastic diffusive conductors. Both Monte Carlo simulations and
analytical approaches are used. Two new phenomena are found: (i) the display
of enhanced shot noise for given energy dependences of the scattering time,
and (ii) the recovery of full shot noise for asymptotic high applied bias.
The first phenomenon is associated with the onset of negative differential
conductivity in energy space that drives the system towards a dynamical
electrical instability in excellent agreement with analytical predictions.
The enhancement is found to be strongly amplified when the dimensionality in
momentum space is lowered from 3 to 2 dimensions. The second phenomenon is
due to the suppression of the effects of long range Coulomb correlations
that takes place when the transit time becomes the shortest time scale in
the system, and is common to both elastic and inelastic nondegenerate
diffusive conductors. These phenomena shed new light in the understanding of
the anomalous behavior of shot noise in mesoscopic conductors, which is a
signature of correlations among different current pulses.
\end{abstract}

\smallskip 
\begin{multicols}{2}

\section{Introduction}

The trend of scaling down the spatial dimensions of conducting samples is a
demanding issue for realizing advanced electron devices and at the same time
a fascinating realm for investigating new physical phenomena. Mesoscopic
systems are presently a subject of intensive research and, being at
intermediate sizes between atomic and micrometric structures, are in the
position to satisfy this issue. In particular, a lot of attention is devoted
to mesoscopic diffusive conductors, which are characterized by a sample
length, $L$, much smaller than the inelastic mean free path, $l_{in}$, and
much longer than the elastic mean free path, $l_{el}$. These systems exhibit
shot noise \cite{Blanter00} when the applied voltage, $V$, is much higher
than the thermal voltage $V_{T}=k_{B}T/q$, where $k_{B}$ is the Boltzmann
constant, $T$ the bath temperature and $q$ the unit charge. Shot noise can
be conveniently studied in terms of the Fano factor $\gamma $ defined as $%
\gamma =S_{I}(0)/(2qI)$, where $S_{I}(0)$ is the low frequency current
spectral density and $I$ the electric current. A value of $\gamma =1$ (full
shot noise) implies the absence of any correlations among different current
pulses, while $\gamma \neq 1$ is a signature of the presence of
correlations. In this latter case, both suppressed ( $\gamma <1$) and
enhanced ($\gamma >1$) shot noise are known to be possible.

In degenerate mesoscopic conductors at $T=0$\cite{Beenaker92,Nagaev92} the
correlations induced by the Pauli exclusion principle give rise to a
universal Fano factor $\gamma =1/3$. By contrast, in nondegenerate
mesoscopic conductors\cite
{Gonzalez98,Gonzalez99a,Gonzalez99b,Beenakker99,Schomerus99a,Schomerus99b,Nagaev99,Gurevich01}
the Fano factor is in general a function of the applied bias and its value
depends on the interplay between the long range Coulomb interaction and the
energy dependence of the elastic scattering time. For the relevant case of a
three dimensional conductor (in momentum space) with an energy dependence of
the scattering time of the form $\tau \propto \varepsilon ^{\alpha }$, Monte
Carlo (MC) simulations\cite{Gonzalez98,Gonzalez99a,Gonzalez99b} have
reported a systematic analysis of the Fano factor in the range of values $%
-1.5<\alpha <1$. Results showed values of $\gamma $ close to $1/3$ for $%
\alpha $ between $0$ and $1$ and $V$ ranging between 40 and 120 $V_{T}$. On
the other hand, the values of $\gamma $ were found to increase
systematically from $1/3$ towards $1$ for values of $\alpha $ decreasing
from $0$ to $-1.5$. The existing analytical theories\cite
{Beenakker99,Schomerus99a,Schomerus99b,Nagaev99,Gurevich01} provide a
reasonable interpretation of these results in the limited range of values $%
-1.5<\alpha <0.5$, being absent \cite{Nagaev99,Gurevich01} or failing
completely \cite{Beenakker99,Schomerus99a,Schomerus99b} outside this range.%
\cite{Gonzalez99b} In this context, two of the Authors \cite{Gomila00} have
recently pointed out the possibility that the above results only provide a
partial description of the shot-noise properties of nondegenerate
conductors. In particular, enhanced shot noise (i.e. $\gamma >1$) was
predicted for given values of $\alpha $, and full shot noise (with $\gamma
=1 $) was expected at asymptotically high bias for $\alpha >-3/2$. We
conclude that the scenario of shot noise in nondegenerate diffusive
conductors, while phenomenological rich of interesting features, is still
incomplete and waiting for a microscopic physical interpretation.

The aim of this work is precisely to address this issue. To this purpose, we
have performed extensive MC simulations for a nondegenerate elastic
diffusive conductor covering values of $\alpha $, applied bias, and
dimensionality that are necessary for a unifying and complete analysis of
the physical problem, and that enable a valuable test of the theoretical
predictions \cite{Gomila00} to be carried out. The paper is organized as
follows. Section \ref{System} describes the physical model and the MC
technique used in the calculations. Section \ref{Theory} summarizes the
theoretical predictions. Section \ref{MonteCarlo} presents the results of MC
simulations proving the possibility for nondegenerate elastic diffusive
conductors to display enhanced as well as full shot noise. The comparison
between elastic and nonelastic transport regimes is also carried out to
outline an interesting anomalous crossover from thermal to full shot noise
exhibited by diffusive conductors. Section \ref{Conclusions} draws the main
conclusions of this work.

\section{Physical model}

\label{System} Following previous works,\cite
{Gonzalez98,Gonzalez99a,Gonzalez99b} nondegenerate diffusive conductors are
modeled as a simple structure consisting of a moderately-doped semiconductor
active region of length $L$ and constant cross-section $A$ sandwiched
between two heavily doped contacts that, by injecting carriers into the
active region, act as ideal thermal reservoirs.\cite{Gonzalez98b} According
with the nondegenerate nature of the carrier injection and the thermal
character of contacts, carriers are injected into the structure following a
Maxwellian velocity distribution at the lattice temperature $T$ and a
Poissonian time statistics, with injection rate $\Gamma =\left( 1/2\right)
n_{c}v_{th}$, where $n_{c}$ is the contact density and $v_{th}=\sqrt{%
2k_{B}T/\pi m}$ the thermal velocity of injected carriers, with $m$ being
the carriers effective mass. Once carriers are injected, their dynamics
(under the action of scattering) is simulated in the active region of the
structure by means of an ensemble MC self-consistently coupled with a
Poisson solver (with constant voltage conditions at the boundaries) to
account for Coulomb correlations.\cite{Gonzalez99a} If not stated otherwise,
we assume that the scattering processes taking place in the active region of
the sample are elastic and isotropic and that the energy dependence of the
scattering time is of the form $\tau (\varepsilon )=\tau _{0}\varepsilon
^{\alpha }$. In principle, the exponent $\alpha $ can take any value,
although only some particular values correspond to well known scattering
mechanisms in a semiconductor model. As examples, $\alpha =-1/2$ corresponds
to scattering with acoustic phonons by deformation potentials, $\alpha =0$
to neutral impurities, $\alpha =1/2$ to acoustic piezoelectric phonons and $%
\alpha =3/2$ to short-range ionized impurity scattering.\cite{seeger85} All
these scattering mechanisms are important at low temperatures, which
represent the typical conditions for an experimental test of theoretical
predictions.

The structure is assumed to be sufficiently thick in transversal dimensions
to allow for a one-dimensional electrostatic treatment. Accordingly, the
Poisson equation is only solved in the direction of the current flow. In
computer simulations one can consider any value of the momentum-space
dimensionality $d$. Thus, even if the physically relevant case is $d=3$, we
will analyze also the case $d=2$, since for this value of $d$ there exist
analytical predictions based on the same assumptions on what concerns the
coupling of the kinetic equation with the Poisson equation.\cite
{Beenakker99,Schomerus99a,Schomerus99b,Nagaev99,Gurevich01,Gomila00} As
indicated in Ref.~[\onlinecite{Schomerus99a}], this case corresponds to a
hypothetical ''flatland'' whose physical realization would be a layered
material with each layer containing a two-dimensional electron gas.

Due to their very high carrier concentration as compared to that of the
active region, contacts are assumed to have no voltage drop inside them. The
applied voltage between the contacts is then considered to be constant in
time. As concerns noise calculations, this situation corresponds to
current-mode operation.\cite{varani94}

In our analysis of shot noise we are mostly interested in determining $%
S_{I}(0)$, which, in MC simulations, is obtained from the time integration
of the autocorrelation function of current fluctuations $C_{I}(t)$. The
calculation of $C_{I}(t)$ is performed from the time series $I(t)$ provided
by the MC simulation. $I(t)$ is the instantaneous steady-state total current
as measured in the outside circuit, which, for the assumed geometry, is also
the same through each cross-sectional area of the device. For the
semiclassical case considered here, the calculation of $I(t)$ is performed
starting from the definition of the microscopic single particle distribution 
\begin{equation}
f({\bf k},{\bf r},t)=\sum_{i=1}^{N(t)}\delta [{\bf k}-{\bf k}_{i}(t)]\delta [%
{\bf r}-{\bf r}_{i}(t)],  \label{fmicro}
\end{equation}
where $N(t)$ is the instantaneous number of electrons inside the sample,
which implies the conduction current density 
\begin{equation}
{\bf j}_{c}({\bf r},t)=q\int {\bf v}f({\bf k},{\bf r},t)d{\bf k}%
=q\sum_{i=1}^{N(t)}{\bf v}_{i}(t)\delta [{\bf r}-{\bf r}_{i}(t)],
\label{jmicro}
\end{equation}
where ${\bf v}_{i}(t)$ is the instantaneous group-velocity determined
through a band structure model in ${\bf k}$-space. We notice that the above
distribution function does contain fluctuations and has the following
properties. Its ensemble average satisfies the Boltzmann equation\cite
{kogan96} while, in the linear approximation of small deviations from its
average stationary-value, it satisfies the Boltzmann-Langevin equation 
\begin{equation}
({\frac{d}{dt}}+S)f({\bf k},{\bf r},t)=y({\bf k},{\bf r},t),
\label{Bolt-Lan}
\end{equation}
with $S$ the linearized scattering operator and $y({\bf k},{\bf r},t)$ the
fluctuating Langevin source related to $S$.\cite{kogan96} From the
solenoidal property of the total current-density, and noticing that the
displacement current is implicitly taken into account by constant-voltage
conditions, it is \cite{varani94} 
\begin{equation}
I(t)={\frac{e}{L}}\sum_{i=1}^{N(t)}v_{i}(t),  \label{Imicro}
\end{equation}
where $v_{i}(t)$ is the instantaneous carrier velocity along the field
direction, known from the MC simulation. We remark that under steady-state
conditions $I(t)$ is a stochastic quantity which enables the calculation of
average values and correlation functions of its fluctuations with respect to
average values. The basic sources of fluctuations in our system come from
the instantaneous changes of carrier velocity due to scattering mechanisms
and of carrier number inside the device due to injection-extraction
processes. Thus, velocity, number and their cross-correlation contributions
to the total current fluctuations are naturally accounted for together with
their coupling with the fluctuating self-consistent field. We remark also
that the ensemble MC simulation, by applying semiclassical dynamics to
describe the motion of each particle between scattering events, is
tantamount to a calculation of the microscopic distribution function $f({\bf %
k},{\bf r},t)$ which, by a suitable discretization of phase space, could be
collected during the simulation. Since two-particle interaction is here
neglected (i.e. the corresponding Boltzmann equation is linear) and the
collision rates are Markovian, the present ensemble MC simulation provides a
numerical solution of the Boltzmann-Langevin equation (\ref{Bolt-Lan}).

Here we are primarily interested in $C_{I}(t)$. This quantity is directly
calculated from the MC by the following procedure. After neglecting an
initial transient, the values of $I(t)$ are recorded in a time grid of step
size $\Delta t$ along the total simulation time $M\Delta t$, with $M$
integer. Then, by defining the time length in which the correlation function
should be calculated as $m\Delta t$, with $m$ integer, $C_{I}(t)$ is
obtained from the algorithm \cite{varani94} 
\begin{eqnarray}
&&C_{I}(j\Delta t)=\overline{I(t^{\prime })I(t^{\prime }+j\Delta t)}-%
\overline{I}^{2}=  \nonumber \\
&=&{\frac{1}{M-m}}\sum_{i=1}^{M-m}I(i\Delta t)I[(i+j)\Delta t]\ -[{\frac{1}{M%
}}\sum_{i=1}^{M}I(i\Delta t)]^{2}
\end{eqnarray}
where the bar denotes time average (ergodicity is implicitly assumed), and $%
j=0,1,...,m$ ; $M>m$. Typically, in our calculations $M>500\times m$, $m>500$
. The corresponding spectral density $S_{I}(f)$ is determined by Fourier
transformation of the $C_{I}(t)$ so obtained.

For the calculations we have used the following parameters: $T=300\ K$,
effective mass $m=0.25\ m_{0}$, with $m_{0}$ the free electron mass, $%
\epsilon =11.7\ \epsilon _{0}$ (with $\epsilon _{0}$ the vacuum
permittivity), $L=200\ nm$ and $n_{c}=4\times 10^{17}\ cm^{-3}$. For these
values $L/L_{D_{c}}=30.9\gg 1$. To test theoretical predictions we have
performed simulations for systems with $d=2,3$ and the following values of $%
\alpha $ (and associated values of $\tau _{0}$) $-2$ ($8.35\times 10^{-16}\ $%
s eV$^{2}$), $-3/2$ ($6.76\times 10^{-16}$ s eV$^{3/2}$), $-1$ ($3.88\times
10^{-16}$ s eV), $-1/2$ ($4.41\times 10^{-16}$ s eV$^{1/2}$) and 0 ($%
2.00\times 10^{-15}\ s$). The values of $\tau _{0}$ are chosen in such a way
to ensure diffusive transport through a sufficiently high number of
scattering events (over $10^{2}$) in the active region. Poisson equation is
solved on a space mesh with $100$ nodes. The number of simulated particles
ranges between 100 and 2000 depending on $\alpha $ and bias conditions. Also
depending on $\alpha $, the time step used to solve Poisson equation ranges
between 2 and 0.1 fs. The numerical uncertainty is confined within 10\% in
average and 20\% at worst.

\section{Theoretical predictions}

\label{Theory} Before Before presenting the Monte Carlo results, we will
briefly summarize and elaborate the theoretical predictions derived in Ref. [%
\onlinecite{Gomila00}] concerning the shot noise properties of nondegenerate
elastic diffusive conductors. In that work, by means of a formal treatment
of the Boltzmann-Langevin-Poisson set of equations it was shown that the
Fano factor for this class of systems can be exactly decomposed into the sum
of three contributions as: 
\begin{equation}
\gamma =\gamma _{in}+\gamma _{\phi }+\gamma _{in,\phi }  \label{FanoSST}
\end{equation}
The first contribution, $\gamma _{in}$, is related to the intrinsic current
fluctuations in the presence of a static non-selfconsistent electric
potential. The second contribution, $\gamma _{\phi }$, is related to the
current fluctuations induced by the fluctuations of the self-consistent
potential. The third contribution, $\gamma _{in,\phi }$, is related to the
cross-correlations between the previous two sources of fluctuations. Under
far from equilibrium conditions (i.e. $V/V_{T}>3$), and for a nondegenerate
conductor, the two following properties are satisfied: $\gamma _{in}=1$, and 
$\gamma _{\phi }\ge 0$. As a consequence of these properties, theory
predicts the following three regimes of shot noise: (a) enhanced shot noise
when $\gamma _{in,\phi }>-\gamma _{\phi }$, (b) full shot noise when $\gamma
_{\phi }=-\gamma _{in,\phi }$, (c) suppressed shot noise when $\gamma
_{in,\phi }<-\gamma _{\phi }$. Concerning enhanced shot noise, Eq.(\ref
{FanoSST}) and the properties $\gamma _{in}=1$, $\gamma _{\phi }\ge 0$ imply
that $\gamma _{in,\phi }>0$ is a sufficient condition to obtain enhanced
shot noise. This condition was found to be physically realized in the
presence of a negative differential conductivity in energy space, i.e. $%
\sigma ^{\prime }(\varepsilon )=d\sigma (\varepsilon )/d\varepsilon <0$,
where the conductivity in energy space is given by\cite{Beenakker99,Gomila00}
\begin{equation}
\sigma (\varepsilon )=\frac{1}{d}q^{2}v^{2}(\varepsilon )\nu (\varepsilon
)\tau (\varepsilon )  \label{sigma}
\end{equation}
with $\nu (\varepsilon )=\Omega m\left( 2m\varepsilon /h^{2}\right) ^{d/2-1}$
the density of states in a $d-$dimensional momentum space, $\varepsilon $
the kinetic energy, $m$ the effective mass, $h$ the Planck constant, and $%
\Omega =2\pi ^{d/2}/\Gamma (d/2)$ the surface area of the unit sphere in $d$
dimensions with $\Gamma (z)$ being the gamma function. Moreover, $%
v(\varepsilon )=\sqrt{2\varepsilon /m}$ is the velocity and $\tau
(\varepsilon )$ the elastic scattering time. Therefore, for $\tau
(\varepsilon )=\tau _{0}\varepsilon ^{\alpha }$ it is 
\begin{equation}
\sigma ^{\prime }(\varepsilon )=C(\alpha +d/2)\varepsilon ^{\alpha +d/2-1}
\end{equation}
where $C$ is a positive constant. We conclude that $\alpha <-d/2$, by
implying $\sigma ^{\prime }(\varepsilon )<0$, represents a sufficient
condition for the presence of enhanced shot noise. This constitutes the main
prediction of the theory\cite{Gomila00} which is here under test.

Concerning full shot noise, Eq. (\ref{FanoSST}) and the property $\gamma
_{in}=1$ imply that $\gamma _{\phi }+\gamma _{in,\phi }=0$ is a sufficient
condition to obtain full shot noise since it corresponds to independent
current pulses through the device. This condition can be realized by the two
alternatives $\gamma _{\phi }=\gamma _{in,\phi }=0$ or $\gamma _{\phi
}=-\gamma _{in,\phi }\neq 0$. The first alternative, $\gamma _{\phi }=\gamma
_{in,\phi }=0$, corresponds to washing out completely the effect of long
range Coulomb correlations. It can be realized under three means: (i) on a
space scale, through the condition that $L<L_{D_{c}}$ where $L_{D_{c}}=\sqrt{%
\epsilon k_{B}T/q^{2}n_{c}}$ is the contact Debye screening length
calculated using the concentration of carriers at the contacts, with $%
\epsilon $ the static dielectric constant of the semiconductor and $n_{c}$
the contact concentration; (ii) on a time scale, through the condition $\tau
_{T}<\tau _{d}$, where $\tau _{T}$ is the transit time and $\tau _{d}$ the
dielectric relaxation time, when $L>L_{D_{c}}$; and (iii) on the energy
dependence of the scattering time, through the condition for the energy
exponent $\alpha =-d/2$ (thus generalizing to any dimensionality the pioneer
3D result of Nagaev\cite{Nagaev99}).

Condition (i) was tested by numerical simulations \cite{Gonzalez99a} that
showed full shot noise for $V/V_{T}>3$ at Debye lengths comparable with the
sample length.

In condition (ii), being the sample length longer than the Debye screening
length, there is room for Coulomb correlations to be effective. However,
since the transit time depends on the bias, the onset of shot noise will
start at voltages in general higher than those satisfying the condition $%
V/V_{T}>3$. The bias value at which shot noise will appear can be roughly
estimated from the equality $\tau _{T}=L^{2}/\mu V_{shot}=\tau _{d}=\epsilon
/q\mu n_{c}$, with $\mu $ being an effective mobility. This gives $%
V_{shot}=qn_{c}L^{2}/\epsilon =(k_{B}T/q)L^{2}/L_{D_{c}}^{2}$, which, since $%
L>L_{D_{c}}$, will lead to $V_{shot}\gg V_{T}$. Therefore, in this case we
expect an anomalous transition between thermal and shot noise for biases $%
V_{T}<V<V_{shot}$. It is worth noting that the expression for $V_{shot}$
just derived coincides with the bias at which $L=L_{s}$, where $L_{s}=\sqrt{%
\epsilon V/qn_{c}}$ is the length used to characterize the space-charge
limited conditions.\cite{Schomerus99a} This makes the condition for the
appearance of full shot noise $\tau _{T}<\tau _{d}$ equivalent to the
condition $L_{s}<L$, thus explaining why the existing theories,\cite
{Beenakker99,Schomerus99a,Schomerus99b,Gurevich01} which apply for $L_{s}>L$
, have failed in predicting the asymptotic presence of full shot noise. In
any case, these theories are expected to remain valid for biases $%
V_{T}<V<V_{shot}$, and their predictions can be taken to explain at least
partially the anomalous crossover between thermal and shot noise where $%
\gamma $ takes values below unity.

In condition (iii), the crossover between thermal and full shot noise is
monotonous and the Fano factor will not take values below one for any bias.

As regards the second alternative to obtain full shot noise, $\gamma _{\phi
}=-\gamma _{in,\phi }\neq 0$, we have not found physically feasible
conditions to accomplish it. From the previous discussion, it follows that
there are two main situations which should be tested by a direct microscopic
simulation, namely: the possibility to observe enhanced shot noise for $%
\alpha <-d/2$ at different dimensionalities, and the possibility to observe
full shot noise for $L>L_{D}$ and $V>V_{shot}$.

\section{Monte Carlo results}

\label{MonteCarlo}

In this section we report the results of MC simulations with the objective
of providing a complete picture of the shot-noise properties of
nondegenerate elastic diffusive conductors and of testing the theoretical
predictions formulated in Sec. \ref{Theory}. Initially we will illustrate
the different conditions under which enhanced, suppressed, and full shot
noise can be obtained in elastic diffusive conductors. Then, the crossover
between thermal and shot noise will be detailed by comparing the noise power
of an elastic diffusive conductor with that of an inelastic one.

\subsection{Enhanced, suppressed and full shot noise}

\begin{figure}[tbp]
\centerline{
\epsfxsize=8cm \epsffile{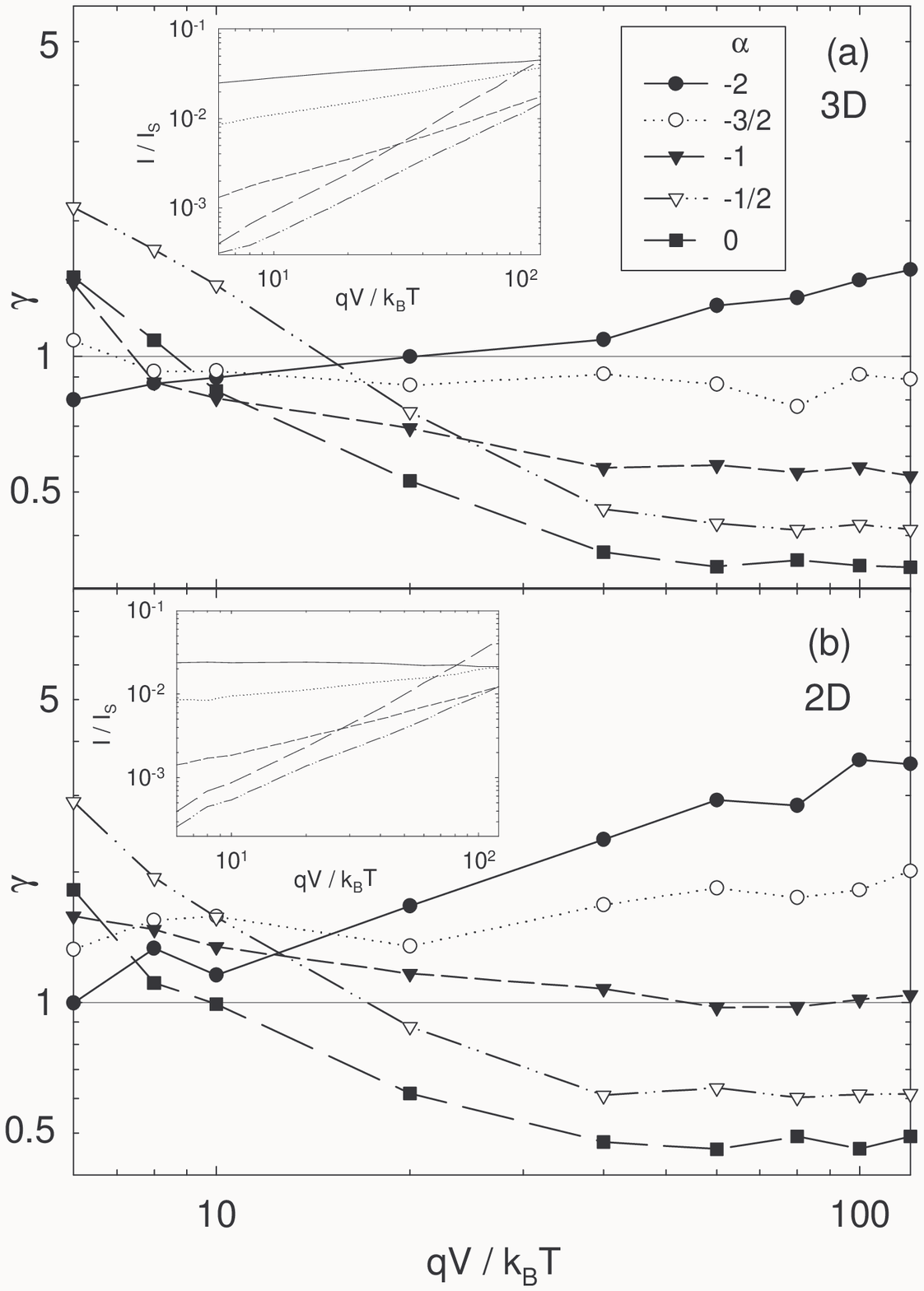}}
\caption{Fano factor as a function of the applied voltage normalized to the
thermal value for $\alpha=-2$,$-3/2$,$-1$ and $0 $. Inset: $I-V$
characteristics with $I$ normalized to the saturation value $I_s$. (a) 3D
system, (b) 2D system.}
\label{FanoEnh}
\end{figure}

Figure \ref{FanoEnh} reports the Fano factor as a function of the normalized
voltage for a major set of simulations performed with different values of
the exponent $\alpha $. Figure 1(a) refers to the 3D case and Fig. 1(b) to
the 2D case. The inset in each figure displays the $I-V$ characteristics in
terms of the saturation current $I_{s}=q\Gamma A=(1/2)qn_{c}v_{th}A$, where $%
A$ is the cross-sectional area of the structure. For voltages below 10 $%
V_{T} $ the conductor approaches thermal noise conditions, which explains a
systematic trend of $\gamma $ to values greater than unity. By contrast, in
the relevant region of voltages from 10 to 100 $V_{T}$ the Fano factor is
found to move from values smaller than 1 to values larger than 1 at
decreasing values of $\alpha $. We note that for values of $\alpha =-2<-3/2$
in 3D ($\alpha =-3/2<-1$ in 2D) the sample displays enhanced shot noise at
increasing voltages, in complete agreement with theoretical predictions for $%
d=2$ and $d=3$. For values of $\alpha >-3/2$ in 3D (and $\alpha >-1$ in 2D)
the system displays suppressed shot noise at increasing voltages, thus
confirming the trend found in previous simulations. \cite{Gonzalez99b} For $%
\alpha =-3/2$ in 3D (and $\alpha =-1$ in 2D) the system is found to display
full shot noise within numerical uncertainty, as expected from theoretical
predictions. Finally, by comparing the general trend of the Fano factor with
the $I-V$ characteristics we find that in passing from suppressed to
enhanced shot noise the conductor exhibits a substantial nonohmic behavior
with a nearly saturation current regime at the largest negative $\alpha $
values.

\begin{figure}[tbp]
\centerline{
\epsfxsize=8cm \epsffile{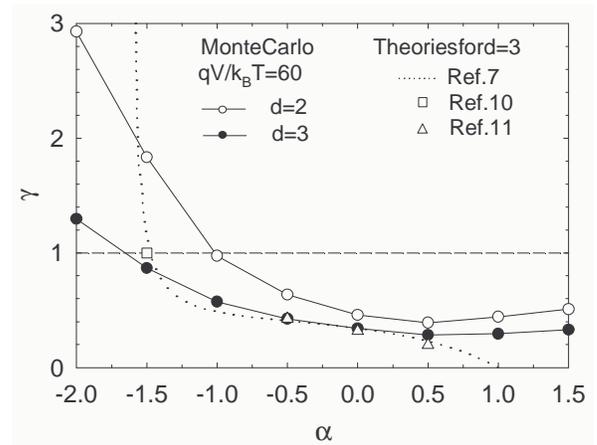}}
\caption{Fano factor as a function of the energy exponent of the elastic
scattering time $\alpha$. Full circles and open circles refer to MC
simulations performed at $V = 60 \ V_T$ for $d=3$ and $d=2$, respectively.
Dotted curve and other symbols refer to analytical theories carried out
under space-charge limited conditions.}
\label{FanoSumary}
\end{figure}

To detail the evolution from suppressed to enhanced shot noise, we have
reported in Fig. \ref{FanoSumary} the value of the Fano factor for the 3D
and 2D cases at $V/V_{T}=60$ for all values of $\alpha $ considered here.
For the purpose of comparison, the figure also reports the theoretical
predictions of Refs. [\onlinecite{Beenakker99,Nagaev99,Gurevich01}] for $d=3$%
. The Fano factor is found to exhibit an asymmetric behavior with $\alpha $,
showing a tendency to saturate, taking suppressed values near to $1/d$, at
positive values of $\alpha $, while increasing systematically towards
enhanced values well above unity at the largest negative values of $\alpha $%
. The quantitative agreement between MC simulations and existing theories is
found to be reasonably good for $-1.5<\alpha <0.5$. These theories are
absent or fail completely outside this range of values. By contrast,
starting from $\alpha \le -d/2$ the simulations evidence an enhanced shot
noise regime which fully validates the theoretical expectations.\cite
{Gomila00}

\begin{figure}[tbp]
\centerline{
\epsfxsize=8cm \epsffile{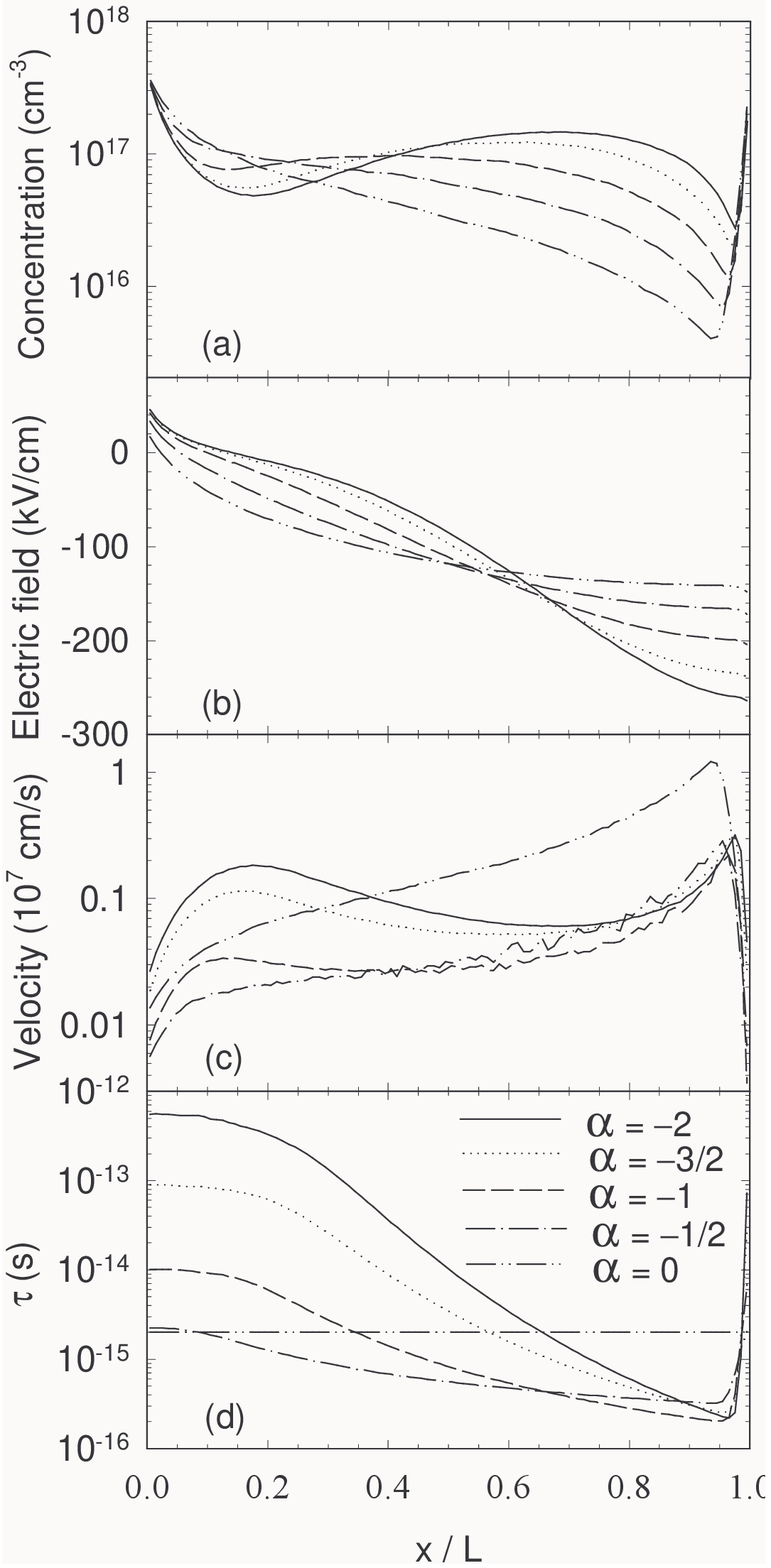}}
\caption{Spatial profiles of (a) carrier concentration, (b) electric field,
(c) average velocity, and (d) relaxation time along the active region of the
sample as obtained by MC calculations for a 3D system, an applied bias of $%
V=80 \ V_T$ and several values of $\alpha$.}
\label{ProfilesEnh}
\end{figure}

To investigate the mechanism responsible for the onset of shot-noise
enhancement, Fig. \ref{ProfilesEnh} reports the spatial profiles of the
relevant average quantities (concentration, electric field, velocity and
scattering time) for an applied voltage $V/V_{T}=80$ and for different
values of $\alpha $ in a 3D system.\cite{note1} From the figure we identify
the signature of a qualitative change in the different profiles for values
of $\alpha $ below or about $-1$. Here, we find the onset of two maxima
(minima) in the velocity (concentration), more pronounced the lower the
value of $\alpha $. This feature determines the presence of two regions
inside the sample characterized by two transport regimes: a quasi-ballistic
one in the region near to the contacts and a diffusive one around the center
of the structure. Furthermore, a wide region of dynamical negative
differential mobility shows up inside the sample, which in turns moves the
structure towards a state of electrical instability, as well known for the
analogous case of Gunn diodes.\cite{Mitin92} These qualitative changes are
accompanied by a systematic increase of the Fano factor, as can be seen in
Figs. \ref{FanoEnh} and \ref{FanoSumary}, what indicates the onset of a
mechanism inducing positive correlations among the fluctuations. Therefore,
the increase of the Fano factor stems as a precursor of the fact that the
system is evolving towards such an electrical instability. The appearance of
a negative differential resistance in the $I-V$ characteristics (more
pronounced in the 2D case), see insets of Fig. \ref{FanoEnh}, further
supports the present interpretation.

\begin{figure}[tbp]
\centerline{
\epsfxsize=8cm \epsffile{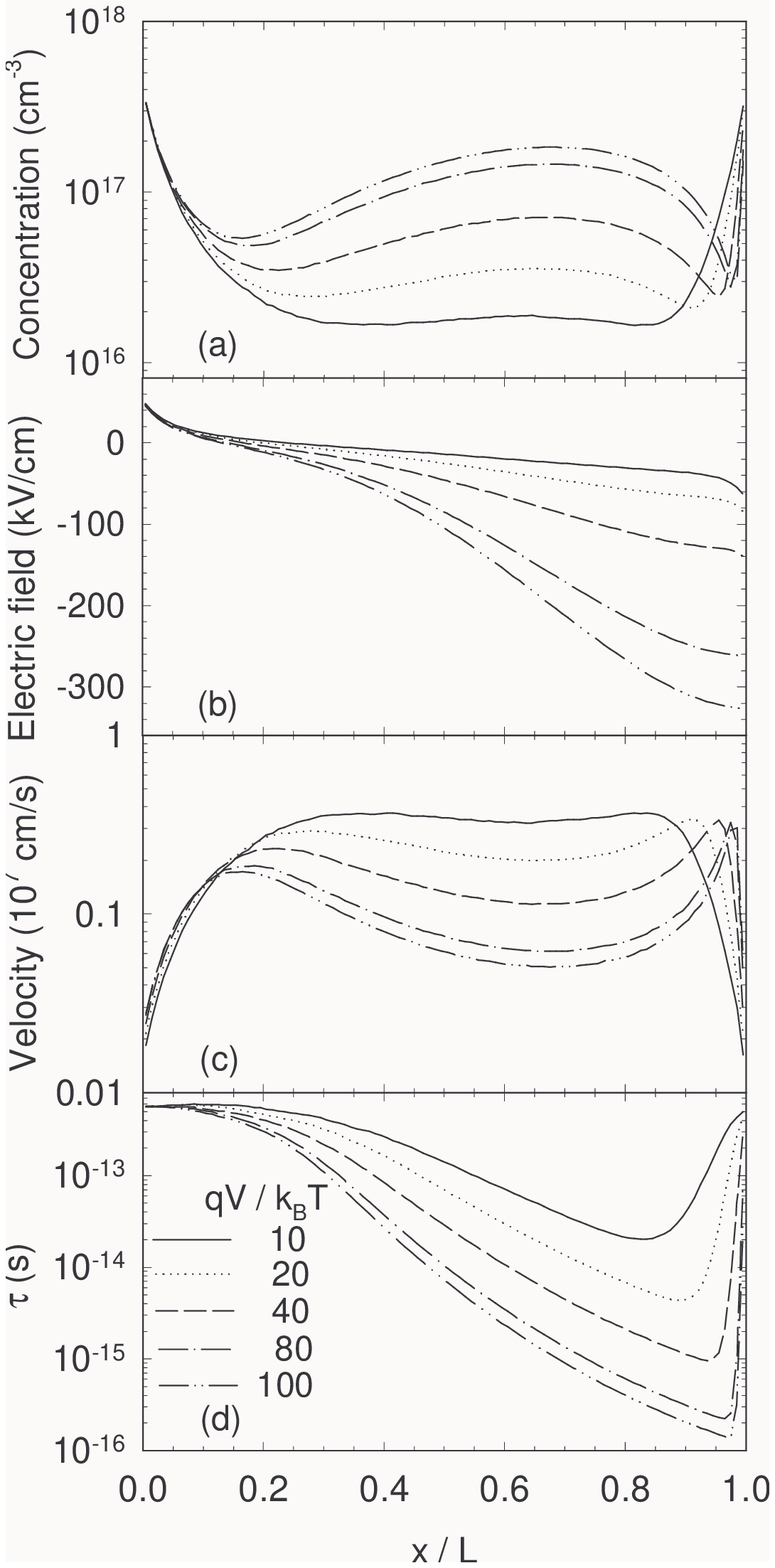}}
\caption{Spatial profiles of (a) carrier concentration, (b) electric field,
(c) average velocity, and (d) relaxation time along the active region of the
sample as obtained by MC calculations for a 3D system, with $\alpha=-2$, and
several values of the applied bias.}
\label{DensityEnh}
\end{figure}

Figure \ref{DensityEnh} reports the spatial profiles of the same relevant
quantities of Fig. \ref{ProfilesEnh} at different applied voltages for the
3D case with $\alpha =-2$. Here the onset of the bimodal profiles of carrier
concentration and velocity at increasing voltages is clearly confirmed, thus
completing the analysis of the microscopic mechanism responsible of the
shot-noise enhancement started with Fig. \ref{ProfilesEnh}.

\begin{figure}[tbp]
\centerline{
\epsfxsize=8cm \epsffile{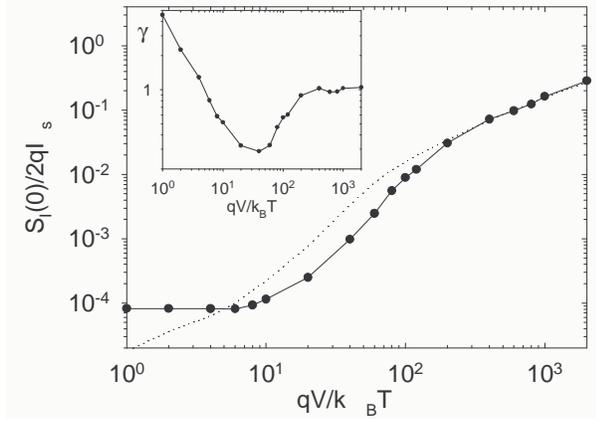}}
\caption{Low frequency current spectral density normalized to $2qI_S$ as a
function of the applied voltage for $\alpha=3/2$ (full circles, continuous
line). The dotted line represents $2qI$. Inset: Fano factor as a function of
the applied voltage for the same case.}
\label{SIa1_5}
\end{figure}

The existing MC simulations \cite{Gonzalez98,Gonzalez99a,Gonzalez99b} and
analytical theories\cite
{Beenakker99,Schomerus99a,Schomerus99b,Nagaev99,Gurevich01} have considered
applied bias up to values high enough to reach the conduction regime known
as space-charge limited conditions and defined by the condition\cite
{Schomerus99a} $L\gg L_{s}\gg L_{D_{c}}$, or equivalently $V_{T}\ll V\ll
V_{shot}$. However, in Sec. \ref{Theory}, starting from suppressed shot
noise we have predicted the possible achievement of full shot noise for bias
values satisfying the condition $V>V_{shot}$, which is in general beyond the
space-charge limited regime. To test this prediction, we have performed MC
simulations for a 3D system with scattering parameters $\alpha =3/2$ and $%
\tau _{0}=8.8\ 10^{-15}$ s eV$^{-3/2}$, at applied voltages sufficiently
high to satisfy $V>V_{shot}$. The results of such simulations are reported
in Fig. \ref{SIa1_5}, which shows the low frequency current spectral density
(continuous curve and full dots) and the current (multiplied by $2q$, dotted
curve) as a function of bias for the simulated structure. For the sake of
completeness, the inset in the same figure shows the corresponding Fano
factor. The simulations prove that at the highest bias the sample indeed
displays full shot noise, thus confirming theoretical expectations. We note
that the values of the applied voltage needed to achieve full shot-noise
conditions are much higher than those reported in Fig. \ref{FanoEnh}.
Focusing on the inset, we observe that the crossover between thermal and
full shot noise does not follow the standard monotonic expression $\gamma
=coth(qV/2k_{B}T)$, but is mediated by a transition region that displays
suppressed shot noise down to a minimum value. We remark that for the
scattering parameters chosen here full shot noise appears at high voltages
for which the $I-V$ characteristic is still far from the current saturation
region. Therefore, this full shot-noise regime can not be related in any way
to the presence of current saturation, as it happens in other systems like
vacuum tubes\cite{Ziel54} or ballistic nondegenerate conductors.\cite
{Gonzalez97,Bulashenko98,Bulashenko00} Rather, it constitutes an intrinsic
property of the sample related to the diffusive transport regime of carriers
(the same phenomenon can be observed in inelastic diffusive nondegenerate
semiconductors, see below and Ref. [\onlinecite{Gomila00b}]). For the
particular value of $\alpha =3/2$ the minimum value of the Fano factor in
the anomalous region of the crossover is around $1/3$. For other values of $%
\alpha \in (-1.5,3/2)$ the value of the minimum is found to lay in the range 
$(1,1/3)$, as shown in Fig. \ref{FanoSumary}. According to these results, we
conclude that for sample lengths longer than the contact Debye screening
length, $L>L_{D_{c}}$, and for $\alpha >-d/2$, a nondegenerate elastic
diffusive conductor exhibits the following three noise regimes: (i) thermal
noise at low bias; (ii) suppressed shot noise, with a minimum value of the
Fano factor determined by the energy dependence of the scattering time ($%
\alpha $), at an intermediate bias range where the sample is under
space-charge limited conditions and; (iii) full shot noise, independently of 
$\alpha $, at the highest bias range.

\subsection{Anomalous crossover between thermal and shot noise}

\begin{figure}[tbp]
\centerline{
\epsfxsize=8cm \epsffile{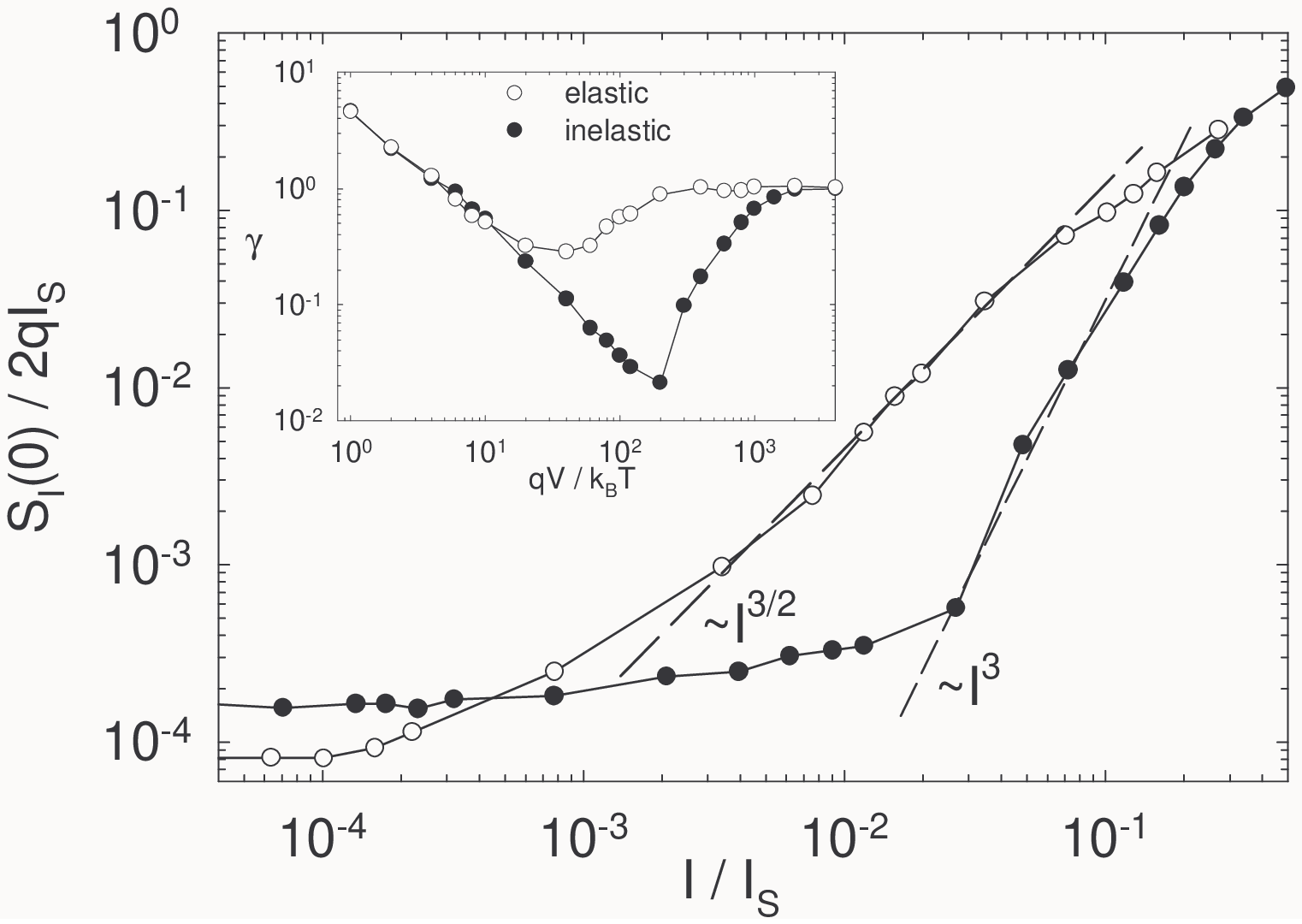}}
\caption{Low frequency current spectral density normalized to $2qI_S$ as a
function of the current normalized to its saturation value for $\alpha=3/2$
in the case of inelastic (full circles) and elastic (open circles)
scattering. Symbols and solid lines refer to MC calculations, and dashed
lines indicate power behaviors on the current in the transition (anomalous)
region defining the crossover between thermal and shot noise. Inset: Fano
factor as a function of the applied voltage for the same cases.}
\label{SIinelas}
\end{figure}

>From the results of the previous section it emerges that the shot-noise
suppression so found can be more generally interpreted in the context of an
anomalous crossover between thermal and shot noise, as introduced in Ref. [~%
\onlinecite{Gomila00b}] for the case of macroscopic diffusive conductors
where scattering is inelastic. To shed more light on this subject, we find
of interest to carry out a comparison between the noise properties of an
elastic nondegenerate conductor with those of an inelastic one. To this
purpose we have performed simulations for a system with the same properties
of that studied above but in the presence of inelastic collisions, using a
scattering model already developed in Ref. [\onlinecite{Gonzalez98}]. The
results are reported in Fig. \ref{SIinelas}, which shows the low frequency
current spectral density for both elastic and inelastic diffusive conductors
as a function of the current flowing through the samples. For the sake of
completeness the inset reports the corresponding Fano factor. In both the
elastic and inelastic cases, at increasing voltages we can distinguish three
different noise regimes, corresponding to thermal, suppressed and full
shot-noise conditions. The degree of suppression of shot noise in the
intermediate region is found to be much more pronounced in the inelastic
case. This fact is more clearly illustrated in the inset of Fig. \ref
{SIinelas}. Remarkably enough, the transition between the thermal and the
full shot-noise regimes in the inelastic case is mediated by a well defined
anomalous crossover with a cubic dependence of the current spectral density
upon the current, in agreement with the results of Ref. [%
\onlinecite{Gomila00b}]. For the elastic case, the crossover displays a
power dependence with current to an exponent of three halves (see Fig. \ref
{SIinelas}), which needs to be further investigated. We conclude that the
anomalous crossover between thermal and full shot noise constitutes a
unifying scenario to interpret shot-noise suppression in nondegenerate
diffusive conductors.

\section{Conclusions}

\label{Conclusions} We have presented a detailed analysis of the noise
properties of nondegenerate elastic diffusive conductors. To this end, we
have performed a complete set of Monte Carlo simulations on the basis of the
indications provided by theoretical predictions. The simulations have
confirmed the existence of enhanced shot noise when the exponent $\alpha $
of the energy dependence of the relaxation time satisfies the condition $%
\alpha <-d/2$, with $d$ the dimensionality of the system in momentum space.
Furthermore, the presence of full shot noise is confirmed under different
conditions, namely: (i) when the length of the sample is shorter than the
contact Debye length, (ii) when $\alpha >-d/2$ and the applied bias is high
enough so that the transit time becomes the shortest time scale of the
system, and (iii) when $\alpha =-d/2$. The presence of enhanced shot noise
for $\alpha <-d/2$ is found to coincide with the analytical condition that
the differential conductivity in energy space becomes negative. The presence
of super-Poissonian values of the Fano factor is here interpreted as
precursor of an electrical instability driven by a region of dynamical
negative differential mobility inside the sample. This instability is
associated with the bimodal spatial profile of the carrier average velocity,
which is generated by the simultaneous presence of two distinct transport
regimes, a ballistic one near to the contacts and a diffusive one inside the
structure. When the sample displays suppressed shot noise the Fano factor
exhibits a minimum whose value depends on $\alpha $. This dependence is in
reasonable agreement with analytical theories developed for space-charge
limited conditions and $d=3$ in the range $-1.5<\alpha <0.5$. Outside this
range analytical theories fail completely providing unphysical values. Since
full shot noise is asymptotically reached at the highest bias, shot-noise
suppression (and the corresponding minimum value of the Fano factor) is more
generally interpreted in terms of an anomalous crossover between thermal and
full shot noise. We believe that this investigation provides a comprehensive
understanding of the noise properties of nondegenerate elastic diffusive
conductors, and sheds new insight into the nonequilibrium noise properties
of mesoscopic conductors.

\section{Acknowledgments}

Partial support from the Ministerio de Ciencia y Tecnolog\'{i}a and FEDER
through the Ramon y Cajal Program and projects Nos TIC2001-1754 and
BFM2001-2159, from the Consejer\'{i}a de Educaci\'{o}n y Cultura de la Junta
de Castilla y Le\'{o}n through the project No. SA057/02, and from the
Italy-Spain Joint Action of the MIUR Italy (Ref. IT109) and MCyT Spain (Ref.
HI2000-0138) is gratefully acknowledged.


\end{multicols}

\end{document}